\documentclass[fleqn,usenatbib]{mnras}

\usepackage{newtxtext,newtxmath}

\usepackage[T1]{fontenc}
\usepackage{ae,aecompl}

\usepackage{graphicx}	
\usepackage{amsmath}	
\usepackage{amssymb}	

\title[Constraining Type I\MakeLowercase{ax} Supernova Progenitor Ages]{Constraining Type I\MakeLowercase{ax} Supernova Progenitor Systems \\ with Stellar Population Aging}

\author[T. Takaro et al.]{
Tyler Takaro,$^{1}$\thanks{E-mail: tktakaro@ucsc.edu}
Ryan J. Foley,$^{1}$
Curtis McCully,$^{2,3}$
Wen-fai Fong,$^{4}$
\newauthor
Saurabh W. Jha,$^{5}$
Gautham Narayan,$^{6}$
Armin Rest,$^{6,7}$
Maximilian Stritzinger,$^{8}$
\newauthor
and Kevin McKinnon$^{1}$
\\
$^{1}$Department of Astronomy and Astrophysics, University of California, Santa Cruz, CA 95064, USA\\
$^{2}$Las Cumbres Observatory Global Telescope Network, Goleta, CA 93117, USA\\
$^{3}$Department of Physics, University of California, Santa Barbara, CA 93106, USA\\
$^{4}$Center for Interdisciplinary Exploration and Research in Astrophysics (CIERA) and \\ Department of Physics and Astronomy, Northwestern University, Evanston, IL 60208, USA\\
$^{5}$Department of Physics and Astronomy, Rutgers, the State University of New Jersey,\\\hspace{.5cm}136 Frelinghuysen Road, Piscataway, NJ 08854, USA\\
$^{6}$Space Telescope Science Institute, 3700 San Martin Dr., Baltimore, MD 21218, USA\\
$^{7}$Department of Physics and Astronomy, Johns Hopkins University, 3400 North Charles Street, Baltimore, MD 21218, USA\\
$^{8}$Department of Physics and Astronomy, Aarhus University, Ny Munkegade 120, DK-8000 Aarhus C, Denmark
}

\date{Accepted XXX. Received YYY; in original form ZZZ}

\pubyear{2018}

\begin{document}
\maketitle

\begin{abstract}
Type Iax supernovae (SNe~Iax) are the most common class of peculiar SNe. While they are thought to be thermonuclear white-dwarf (WD) SNe, SNe~Iax are observationally similar to, but distinct from SNe~Ia. Unlike SNe~Ia, where roughly 30\% occur in early-type galaxies, only one SN~Iax has been discovered in an early-type galaxy, suggesting a relatively short delay time and a distinct progenitor system. Furthermore, one SN~Iax progenitor system has been detected in pre-explosion images with its properties consistent with either of two models: a short-lived (<100 Myr) progenitor system consisting of a WD primary and a He-star companion, or a singular Wolf-Rayet progenitor star. Using deep \textit{Hubble Space Telescope} images of nine nearby SN~Iax host galaxies, we measure the properties of stars within 200 pc of the SN position.  The ages of local stars, some of which formed with the SN progenitor system, can constrain the time between star formation and SN, known as the delay time.  We compare the local stellar properties to synthetic photometry of single-stellar populations, fitting to a range of possible delay times for each SN.  With this sample, we uniquely constrain the delay-time distribution for SNe~Iax, with a median and $1-\sigma$ confidence interval delay time of $63_{- 15}^{+ 58} \times 10^{6}$ years. The measured delay-time distribution provides an excellent constraint on the progenitor system for the class, indicating a preference for a WD progenitor system over a Wolf-Rayet progenitor star.
\end{abstract}




\section{Introduction}

Type Iax supernovae (SNe~Iax) are a class of peculiar SNe which share some characteristics with Type Ia SNe, but appear to be physically distinct \citep{fol13}. SNe~Iax are also known as SN 2002cx-like SNe, after the prototypical object of the class \citep{li03, jha06, jha17}. These SNe are principally characterized by their low luminosities (as compared to SNe~Ia), low photospheric velocities, a lack of a secondary maximum in near-IR bands \citep{li03}, and their unique late-time spectra \citep[e.g.,][]{jha06, fol16}. In contrast to SNe~Ia, SNe~Iax reach their peak brightness in optical bands before they do in the near-IR.

SNe~Iax are the most common type of peculiar SN, occurring at roughly 30\% the rate of SNe~Ia \citep{fol13, miller17}. This, along with their observational similarities with SNe~Ia, means that investigating the physical properties of these supernovae is extremely helpful to understanding SNe~Ia. The comparably low photospheric velocities of SNe~Iax makes line identification easier, allowing for precise measurements of the explosion in these events, which can help us to understand the physics involved in both SNe~Iax and Ia. This will be crucial as we look to improve our knowledge of SNe~Ia in order to continue to perform precision cosmology. Despite decades of study of SNe~Ia, the identity of the binary companion to the SN~Ia progenitor is still unknown. By first learning about the progenitor system for SNe~Iax, we may be able to shed some light on this long-standing problem.

As compared to typical SNe~Ia, SNe~Iax are also inferred to have much lower explosion energies \citep{bran04}, and lower ejecta masses \citep{fol13} (though this has been debated for some members of the class, e.g., \citealt{sahu08}, \citealt{stritz15}, \citealt{yama15}). Additionally, SNe~Iax tend to have a layered structure in their ejecta, similar to SNe~Ia. In contrast to SNe~Ia, however, there is evidence of Ni mixing in at least two SNe~Iax, further from the center than in standard SNe~Ia \citep{jha06, phil07, stritz15}. Finally, at least six SNe~Iax show cobalt in their near-IR spectra \citep{krom13, stritz14, stritz15, toma16}. Taken together, these properties suggest that SNe~Iax may be the result of the partial deflagration of a white dwarf (WD) \citep[e.g.][]{fol10a, krom13, fink14, mag16}.

SNe~Iax host galaxies are also markedly different from SNe~Ia host galaxies. While the SNe are nearly always found in late-type galaxies, there is one example (SN 2008ge) of a SN~Iax in an S0 galaxy \citep{fol10b}. \citet{lym13} and \citet{lym18} find evidence of star forming regions near the sites of most SNe~Iax, suggesting a young progenitor age at the time of explosion for these events. Additionally, SNe~Iax seem to preferentially occur in metal-poor regions of their host galaxies \citep{lym18}. This is quite different from the metallicity of the explosion sites for SNe~Ia, instead matching quite closely with the sites of SNe Ib, Ic, II, and IIb. Though the metallicities of the explosion sites for SNe~Iax and core-collapse SNe are similar, they appear quite different from those of low-redshift long gamma-ray bursts (LGRBs; \citealt{lev10}). This is particularly interesting, as the most popular core collapse model for SNe~Iax is the ``fallback SN'' model \citep{val09, mor10}, which are theorized to occur alongside LGRBs \citep{heg03, del06}. Thus the divergent metallicity distributions of SNe~Iax and LGRBs suggest that SNe~Iax may not be ``fallback'' core-collapse SNe.

The physical origin of SNe~Iax is still uncertain, but the leading models suggest that they are the result of binary interactions between a carbon-oxygen WD and a He-star companion \citep{fol13, jha17}. These interactions seem to result in an incomplete deflagration (sub-sonic nuclear burning, \citealt{phil07}) of the WD which may not completely unbind the star \citep[e.g.,][]{jord12, krom13, fink14, long14, mag16}. The low luminosities, velocities, and ejecta masses measured from these events together give strong indirect evidence for a WD progenitor system \citep{fol09, fol10a}. Additionally, the Ni mixing observed in some SNe~Iax ejecta can be most easily explained by a turbulent deflagration in a WD \citep{rop05}. Despite the mounting evidence for this model, there remain several unresolved issues. Specifically, the incomplete deflagration model struggles to reproduce the observed lower luminosity SNe~Iax, such as SNe 2008ha and 2010ae \citep{krom13, fink14} It has also been argued \citep{krom15} that SN 2008ha spectra are indicative of poorly mixed ejecta, in contrast to brighter SNe~Iax. Well-mixed ejecta are a hallmark of partial deflagration models, making SN 2008ha particularly confusing. Additionally, at least two SNe~Iax (SNe 2004cs and 2007J) have He lines in their spectra, which is difficult to reproduce with the WD and He-star binary model \citep{fink14}.

For the WD He-star binary progenitor model for SNe~Iax, it is expected that the circumstellar environment will be enriched by pre-SN mass loss, either from winds from the He donor star, or from non-conservative mass transfer. Once the WD explodes, the interaction between the blast wave and the circumstellar material should power X-ray and radio emission \citep{chev06, imm06, rus12, marg12, marg14}. Though X-ray emission has not yet been detected from any SNe~Iax, \citet{liu15} use X-ray upper limits from seven SNe~Iax \citep{rus12} to constrain their pre-SN mass loss rates. In comparing theoretical pre-SN mass loss rates to the observed X-ray upper limits, they find broad agreement across a variety of models and observations.

The WD He-star binary model for SNe~Iax has also been tested with pre-explosion \textit{Hubble Space Telescope} (\textit{HST}) imaging of the locations of SN 2008ge \citep{fol10b}, SN 2008ha \citep{fol14}, SN 2012Z \citep{mccul14b}, and SN 2014dt \citep{fol15}. These works use precision astrometry to align pre-explosion images with ground-based images of the SN. The authors then perform photometry on the pre-explosion images at the site of the explosion to probe the progenitor system just before explosion. In doing so, they find an upper bound on the progenitor mass for SN 2008ge (assuming a massive star progenitor), find an upper bound on the age of the SN 2008ha progenitor system, and resolve the likely progenitor system of SN 2012Z. \citet{fol15} find no progenitor system for SN 2014dt down to quite deep limits ($3-\sigma$ limits of $M_{F438W} > - 5.0$ mag and $M_{F814W} > - 5.9$ mag), making the interpretation of SN 2014dt as a core-collapse event less likely. While these limits rule out most Wolf-Rayet star models, there are some Wolf-Rayet star models which remain below the detection limits. In all four cases, the results show consistency with the WD and He-star model, though the large photometric errors involved prevent them from ruling out other models with high certainty.

The likelihood of observing a WD progenitor in pre-explosion imaging of SN~Iax explosion sites is quite low, due to the inherent low luminosity of WDs as compared to their mass donor companion. However, analyzing the host environments of SNe (in pre or post-explosion imaging) has been quite successful in extracting information on the progenitors of Type Ib, Type Ic, and Type II SNe \citep[e.g.,][]{maiz04, bad09, kunc13a, kunc13b}. One such method uses resolved stellar populations near the SNe, fit to theoretical isochrones (lines of constant age on a color-magnitude diagram) to measure the age of the SN progenitor \citep[e.g.,][]{gog09, murph11, jenn12, wil14, jenn14}. In this method, it is assumed that the resolved stellar population and the progenitor itself formed nearly simultaneously, in a single burst of star formation. This method of aging stellar populations to measure time between star formation and explosion -- known as the delay-time -- for SNe has only been effective for SNe with relatively short delay times, as typical open cluster velocity dispersions are large enough that stellar positions are no longer correlated after a few hundred million years \citep{bast06, lada10}. Every model for SNe~Iax predicts a delay-time distribution (DTD) with non-zero probability down to below 100 Myrs, so it is reasonable to apply this method to measure SNe~Iax delay times.

In the WD progenitor model for SNe~Iax, explosions occur at the Chandrasekhar mass, and the quickest binary channel for a carbon/oxygen (C/O) WD is to accrete helium from a He star companion \citep{hach99, post14}. The stable mass transfer rate can be high for helium accretion, and \citet{cla14} show that this channel dominates the thermonuclear SN rate between 40 Myr and 200 Myr (above which traditional SNe~Ia dominate). This has been observed in several binary population synthesis studies \citep{rui09, rui11, wang09a, wang09b, meng10, pier14, liu15}. \citet{liu10} present a model (originally intended to explain a different kind of system) that begins with a 7 M$_\odot$ + 4 M$_\odot$ binary system that undergoes two phases of mass transfer and common envelope evolution, before resulting in a 1 M$_\odot$ C/O WD + 2 M$_\odot$ He star. As the He star evolves, it again fills its Roche lobe and begins stable mass transfer onto the white dwarf that could lead to the SN~Iax. The high accretion rate involved in this process mean that total delay time is dominated by the stellar evolution timescale of the secondary, giving short expected delay times for these models.

In this work, we employ Bayesian Monte Carlo methods to fit isochrones to the stellar populations around SNe~Iax, in order to accurately measure the DTD of this class of SNe. With a sample of nine SNe~Iax, we constrain the ages of these systems at the time of explosion.

\citet{li18} examine the location of SNe~Iax in their host galaxies, and find that they tend to occur at much larger projected radii as compared to SNe~Ia. The authors perform a Kolmogorov-Smirnov test to compare the fractional host galaxy fluxes at the explosion sites for SNe~Iax, SNe~Ia, and SNe Ib/Ic. They find strong evidence that SNe~Ia and SNe~Iax are drawn from different populations, while also finding that they cannot reject the hypothesis that SNe~Iax and SNe Ib/Ic are drawn from the same fractional host galaxy flux distribution. Taken together with the metallicity information, this suggests that the delay-time for SNe~Iax might be much more similar to those of core-collapse SNe than the delay-times measured for SNe~Ia.

This paper is structured in the following way. In Section~\ref{sec:obs}, we describe the observations used in this study, and the methods to extract photometry of the stellar populations near each SN. In Section~\ref{sec:iso}, we detail the Bayesian Monte Carlo methods used to probabilistically determine the ages of these stellar populations. We discuss the particulars of each SN individually in Section~\ref{sec:indv}, describing the priors used in the fit and the resulting posterior distributions. In Section~\ref{sec:disc}, we discuss the overall properties of the measured DTDs for this sample of SNe~Iax, and describe how our results compare to theoretical models for this class of SN.

\section{Observations and Data Reduction}\label{sec:obs}

\subsection{Sample Selection}
The distance to a host galaxy is the key parameter which restricts our sample. Because the methods we will employ depend on analyzing a population of well resolved stars in the direct vicinity of each SN under consideration, only relatively nearby targets are suitable. Because of this limitation, our 2013 \textit{HST} program (GO-12999) observed SNe 2008ge \citep{pig08}, 2008ha \citep{fol09}, 2010ae \citep{pig10}, and 2010el \citep{mon10}. At the time our \textit{HST} program was executed, these were the four closest known SNe~Iax. They were selected as a sample to demonstrate that stellar population fitting could be a reliable method to probe the delay-time of SNe~Iax. These four objects were observed with the Advanced Camera for Surveys Wide Field Channel (ACS/WFC), with the F435W, F555W, F625W, and F814W filters. These bands were chosen because they provide good colors for observing broad spectral types, and their spectral range enables one to correct for extinction due to dust in the local environment.

In the time since 2013, we have searched the \textit{HST} archive for other SNe~Iax with high quality data (wide spectral range and long exposures) of their stellar neighborhoods. This search added SNe 2008A (GO-11590: Jha, S.), 2011ay (GO-15166: Filippenko, A.), 2012Z (GO-10497: Riess, A.; GO-10711: Noll, K.; GO-10802: Riess, A.; and GO-13757: Jha, S.), 2014ck (GO-13029: Filippenko, A.), and 2014dt (GO-13683: Van Dyk, S.; GO-14779: Graham, M.) to our sample. Two of these objects have ACS/WFC imaging, with SN 2012Z imaged in the F435W, F555W, and F814W bands, and SN 2008A imaged in the F555W, F625W, and F775W bands. SN 2011ay and SN 2014ck each have Wide Field Camera 3 (WFC3) imaging, in the F555W and F814W bands, and the F625W and F814W bands, respectively. Additional imaging of SNe 2010el was added in this search (GO-13364: Calzetti, D.; GO-13816: Bentz, M.; GO-14668: Filippenko, A.; and GO-15133: Erwin, P.). Finally, our 2015 follow-up \textit{HST} program for SN 2008ha (GO-14244: Foley, R.) provided additional imaging in the F435W and F814W filters.

Our complete sample of SNe~Iax contains all observed SNe~Iax within 35 Mpc, for which \textit{HST} host galaxy imaging is deep enough to find strong constraints on progenitor age. In addition, two more distant SNe (2008A and 2011ay) are included in the sample, thanks to the abundance of \textit{HST} imaging of their host galaxies. SNe 2008A, 2008ge, 2008ha, 2010ae, 2010el, 2011ay, 2012Z, 2014ck, 2014dt compose our sample. The complete list of data can be found in Table~\ref{table:results1}. Of these, previous works have found constraints on the progenitors of SNe 2008ge, 2008ha, and 2012Z. These constraints were achieved through pre-explosion \textit{HST} imaging of the sites of SNe 2008ge \citep{fol10b} and 2012Z \citep{mccul14b}, and post-explosion imaging of the site of SN 2008ha \citep{fol14}, with claims of detections of either the remnant, or the probable donor star for both SN 2008ha and SN 2012Z.

Note that for host galaxy distances which are measured using redshift, we assume $H_0 = 73.24 \pm 1.74$ km Mpc$^{-1}$ s$^{-1}$ \citep{riess16}. The distances to each object, as well as the methods with which these distance were measured, can be found in Section~\ref{sec:indv}. The  Additionally, Milky Way reddening is assumed to follow the extinction maps of \citet{schl11}, using $R_v = 3.1$ \citep{card89}. This reddening is applied to all simulated data to match the observations.

\subsection{Data Reduction}
To build catalogs from the \textit{HST} observations, we use a custom pipeline written primarily in {\tt Python}\footnote{\url{https://github.com/cmccully/snhst}}. The pipeline initially registers the astrometry to a ground-based image which has a wider field of view than the \textit{HST} images. The World Coordinate System (WCS) from the ground-based image is considered to be the global astrometric solution. This stage typically produces a precision of $\sim0.05\arcsec$ which corresponds to roughly one \textit{HST} pixel for ACS.

After solving for the global WCS, we refine the registration between individual exposures, starting from the flat-fielded frames from the MAST archive\footnote{\url{https://archive.stsci.edu/}}. For ACS and WFC3, we use the FLC frames that have been corrected for charge-transfer inefficiency (CTI) using the pixel-based method \citep{Anderson10}. Cosmic rays are rejected from the individual frames using {\tt Astro-SCRAPPY}\footnote{\url{https://github.com/astropy/astroscrappy}\label{foot1}} before the registration process to alleviate false positives in the catalog matches. After doing a coarse registration between \textit{HST} visits by hand, the pipeline uses TweakShifts from {\tt Drizzlepac}\footnote{\url{https://drizzlepac.stsci.edu}} \citep{driz} to refine the offsets between frames. 

Once the registration is completed, we combine the exposures using {\tt Astrodrizzle} adopting standard values for the parameters$^{\ref{foot1}}$. We include the cosmic-ray step in {\tt Astrodrizzle} to do final cosmic-ray rejection. For cases that we have more than 4 individual exposures in the same filter for a target, we subsample the pixel grid and decrease the drizzle pixel coverage fraction to 0.8. 

To build the final catalogs, we run {\tt DOLPHOT}, a modified version of HSTPhot \citep{Dolphin00}, using the drizzled image as a coordinate reference. {\tt DOLPHOT} runs on the individual flat fielded frames and stacks the photometry to produce the final catalog, including all point sources across the entirety of each image. We again use the CTI-corrected FLC frames for ACS and WFC3/UVIS. For WFPC2, we use the CTI correction built into {\tt DOLPHOT}. {\tt DOLPHOT} includes its own image registration stage which generally produces a scatter of $\sim 0.01\farcs$ Finally, we inject artificial stars using {\tt DOLPHOT} to estimate the brightness limit of our images.

\subsection{Star Selection}
Using the {\tt DOLPHOT} output, we apply the recommended cuts in sharpness (<0.3), roundness (<1), and crowding (<0.1), along with a minimum signal-to-noise ratio of 3.5. Though this removes most non-stellar detections, there remain a number of clusters and bright clumps of gas which are selected as stars by {\tt DOLPHOT}. As such, for each region surrounding a SN, every detected source within the region is checked by eye to ensure that it has a point spread function indicative of a single stellar source. This allows us to remove gas clumps from our catalogues.

As the human eye struggles to distinguish distant clusters from stars in the galaxy under examination, another method is used to make this distinction. Aperture photometry is performed using IRAF in order to determine a concentration parameter for each source \citep{cha10}. The concentration parameter gives the difference in measured F555W (or a similar band) magnitude between when a 3 pixel aperture is used, and when a .5 pixel aperture is used. By separating on this concentration parameter, as detailed in \citet{cha10}, we remove the extended sources from our sample, leaving us with only sources that have a high likelihood of being stars.

\section{Isochrone Fitting Method}\label{sec:iso}

In order to determine the age of a supernova progenitor in in post-explosion host galaxy images, we assume a star formation history (SFH) characterized by a dominant star formation event for the region in which the progenitor system formed. We also assume a typical scale within which we can reasonably expect nearly every star examined to have a shared SFH. Past work has indicated that this radius is anywhere from 50 pc to 200 pc \citep{bast06, eld11, wil14, fol14, maund16, maund17}. To account for this uncertainty, we consider all stars within a 200 pc projected on-sky radius of the SN position, weighting the stars according to their probability of being associated with the SN.

To estimate the probability of association as a function of on-sky distance, we build a probability distribution as follows. We first assume a flat initial distribution of stars within 100 pc of the SN, to represent the cluster at the time of formation. We then assume a velocity dispersion of $0.65$ km/s \citep{gel09}, multiplied by the age of the cluster (in this case the delay-time being tested), to model the cluster spreading out over time. To ``apply'' this spreading effect to the initial flat distribution, we convolve the two distributions. This convolution of the two distributions gives us a rough probability of association with the SN as a function of projected physical distance. This probability function is used to weight the stars within 200 pc, with the higher probability stars receiving a higher weight in the fitting scheme. We fit a single stellar population model to each of these populations.

\citet{gel09} quotes a velocity dispersion for clusters of $0.65 \pm 0.10$ km/s. To account for this uncertainty, we run our complete analysis using dispersions of $0.55$ km/s, $0.65$ km/s, and $0.75$ km/s. Each of these choices ultimately leaves the final PDF in delay time largely unchanged. For this reason, we use the median value of $0.65$ km/s in all the analyses that follow.

\citet{maund17} uses a similar technique to age the stellar population around the positions of 12 Type IIP SNe. In that paper, however, they fit to multiple stellar populations for each SN, allowing for multiple bursts of star formation in the SFH. In this study, we choose to only fit a single stellar population around each SN, as the errors in magnitude and color for our detected stars are large enough that fitting to multiple populations would result in overfitting.

\subsection{Color Magnitude Diagrams}
To date the regions around the SNe, we use the magnitudes and colors derived using {\tt DOLPHOT} to place each star on a color magnitude diagram (CMD). We then overplot the {\tt MIST} synthetic photometry isochrones \citep{pax11, pax13, pax15, dot16, cho16}, corrected for distance, metallicity, and extinction. We test isochrones in the age range $10^{6.5}$ to $10^{8.5}$ years. For metallicity and extinction, we test the $2$-$\sigma$ range from our priors. One sigma errors in distance and internal (host galaxy) reddening are propagated through to color and magnitude, added in the appropriate way and displayed in CMDs as shaded regions.

\begin{table*}
    \setlength\tabcolsep{3pt}
    \centering
    \caption{ The list of observations used in this study. The data can be found here: \href{http://dx.doi.org/10.17909/t9-qr61-xb59}{http://dx.doi.org/10.17909/t9-qr61-xb59} \label{table:results1}}
    \begin{tabular}{@{\hspace{-.35cm}}cccccc}
    \hline 
    SN & Instrument & Filters & Number of Exposures \hspace{-20pt} & Total Exposure & Date of Observation \\
    Name & & & per Filter & Lengths (sec) & (UTC) \\
    \hline
    \small
    \vspace{0.15cm} 2008A & ACS/WFC & F555W, F625W, F775W & 6, 6, 4 & 3750, 3530, 2484 & 08/18/2009 \\
    \vspace{0.15cm}     "    & WFC3/IR & F110W & 12 & 8336 & 08/18/2009 \\
    \vspace{0.15cm} 2008ge & ACS/WFC & F435W, F555W, F625W, F814W & 2 & 1168, 768, 844, 1244 & 10/26/2012 \\ 
    \vspace{0.15cm} 2008ha & ACS/WFC & F435W, F555W, F625W, F814W & 2 & 9068, 764, 840, 12058 & 01/02/2013, 12/30/2015 \\ 
    \vspace{0.15cm} 2010ae & ACS/WFC & F435W, F555W, F625W, F814W & 2 & 1402, 1002, 1078, 1478 & 05/23/2014 \\ 
    \vspace{0.15cm} 2010el & ACS/WFC & F435W, F555W, F625W, F814W & 2 & 1168, 768, 844, 1244 & 05/23/2013 \\
    \vspace{0.15cm}     "     & WFC3/UVIS & F275W, F336W, F438W, F555W, F814W & 3, 3, 3, 5, 5 & 2382, 1119, 965, 1853, 1769 & 09/2013, 07/2015, 08/2017 \\ 
    \vspace{0.15cm} 2011ay & WFC3/UVIS & F555W, F814W & 2 & 780, 780 & 01/20/2018 \\ 
    \vspace{0.15cm} 2012Z & ACS/WFC & F435W, F555W, F814W & 8, 10, 10 & 9624, 12642, 12868 & 09/2006 \\ 
    \vspace{0.15cm} 2014ck & WFC3/UVIS & F625W, F814W & 2 & 510, 680 & 02/22/2013 \\
    \vspace{0.15cm} 2014dt & WFC3/UVIS & F275W, F438W & 2, 20 & 858, 400 & 02/28/2017, 11/18/2014 \\
    \hline
    \end{tabular}
\end{table*}

\subsection{Isochrones}
In order to fit the synthetic photometry isochrones to the detected stars accurately and with well understood errors, a hierarchical Bayesian framework is used to create a new statistic to measure goodness of fit\footnote{\url{https://github.com/TTakaro/Type-Iax-HST}}. Monte Carlo methods are then used to translate this statistic into probability distributions. In analogy to a Chi-squared fit, this statistic (which we will hereafter call IGoF for ``Isochrone Goodness of Fit") measures the minimized difference in magnitude in each filter between the stars detected near the SN and the isochrone being tested, with the magnitude difference in each filter then summed in quadrature. Summing over each detected star, IGoF gives a total distance from the stars to the isochrone in n-dimensional magnitude space, where n is the number of filters. As described in the previous section, each star is weighted in IGoF according to its projected physical distance from the SN, and thus its likelihood of forming coincident with the SN progenitor system. This statistic was selected in part because of its qualitative similarity to a Chi-squared fit, allowing for relatively simple analysis.

The equation for IGoF is shown below.
\begin{equation}
    \mathrm{IGoF} = \frac{\sqrt{\sum_{i} \mathrm{PhysDist}_{i} \cdot min(\sum_{k} (m_{ik} - m_{\mathrm{Iso.}, k})^{2})}}{\sum_{i} \mathrm{PhysDist}_{i}},
\end{equation}
    
\begin{align}
    \mathrm{PhysDist} &= \mathrm{Flat}(min=0 \mathrm{pc}, max=100 \mathrm{pc}) \circledast \\ &  \mathrm{Normal}(\mu=0, \sigma=.65 \mathrm{\frac{km}{s}} \cdot \mathrm{Age}), \notag
\end{align}
where $i$ specifies a given star, $k$ specifies a given filter, $m_{ik}$ is the magnitude of a given star in a filter, $m_{\mathrm{Iso}, jk}$ is the magnitude drawn from the isochrone, adjusted for distance, milky way reddening, and host galaxy reddening. $\mathrm{PhysDist}_{i}$ is the weight given to the star $i$ according to its physical distance from the SN. As detailed in section~\ref{sec:iso}, this weight is the convolution of a flat distribution, and a normal distribution, evaluated at the on-sky distance measured between star $i$ and the SN. Because a cluster of stars disperse as the cluster ages, this weighting function will have a different shape at different ages.

To translate the measured value of IGoF for an isochrone into a relative probability that the stars in the data match that isochrone, careful forward modeling is required. A set of artificial stars equal in number to the detected stars are generated from each isochrone, drawn from normal distributions in flux space using the characteristic flux error from the detected stars in the host galaxy image. These stars are given radial distances on the sky generated from the radial distribution used to weight stars in the IGoF. An IGoF value is then measured for the set of artificial stars. This process is repeated 5000 times for each isochrone, until the histogram of IGoF values converges to a probability distribution. This probability density function (PDF) gives a probability of measuring a value of IGoF, given the chosen age, metallicity, and host galaxy reddening associated with the isochrone. Using this fit PDF, the values of IGoF measured from the data are translated into values of relative probability for each isochrone. These relative probabilities are then normalized to determine a relative probability distribution in age, metallicity and host galaxy extinction for each SN. We then marginalize over metallicity and host galaxy extinction in order to extract the one dimensional age distributions for each object.

Further details of our method, including the likelihood function used in the isochrone fitting scheme are shown in Appendix~\ref{app}.

\subsection{False Star Tests and Photometric Completeness}
To determine the completeness of our photometry, we perform false star recovery tests using the false star tool in {\tt DOLPHOT}. For host galaxy image, we insert 50,000 stars with magnitudes covering the whole range of measured magnitudes, and with x-y positions covering the entirety of the region around the SN. Using the results of the false star runs in {\tt DOLPHOT} and binning with 0.1 mag resolution, we calculate a recovery fraction as a function of magnitude for each host galaxy image. We then apply this recovery fraction to the artificial stars generated while calculating IGoF in the process mentioned above.

In several of the imaged SN host galaxies, no stars are detected within 200 pc of the SN position. In order to get an upper bound on the age of the progenitor star for these SNe, we use the recovery fraction calculated using false star testing. From this recovery fraction function, we determine a limiting magnitude of detection in the image, by requiring that 90\% of inserted stars be recovered for each magnitude bin above our limiting magnitude. We then simulate the effects of a $ 50 M_{\odot}$ open cluster (\citealt{lada10} find $50 M_{\odot}$ to be the most common open cluster mass) at the position of the SN under consideration, drawing stars from a Kroupa Initial Mass Function (IMF) \citep{krou01} until we reach the total cluster mass. If any of these drawn stars have brightnesses above our limiting magnitude given the isochrone we are considering, we rule out the associated age, metallicity, and reddening combination for this SN. By performing this analysis for each isochrone, we establish a lower bound on the age of each SN progenitor system for which we detect no nearby stars (SNe 2008A, 2008ge, and 2011ay).

\section{Analysis of Individual Objects}\label{sec:indv}
The objects in our sample and the associated data used for this study are listed in Table~\ref{table:results1}.

\subsection{SN 2008A}
SN 2008A was discovered in NGC 624 \citep{nak08}, at an estimated Tully-Fisher distance of $51.5 \pm 11.0$ Mpc \citep{theu07}. For this object, we assume no host galaxy extinction, as the SN is located on the fringes of it's host galaxy \citep{lym18}. The metallicity for this object is difficult to measure \citep{lym18}, so we use solar metallicity as our prior \citep{asp09}. For additional analysis of SN 2008A, see e.g., \citet{miln10}, \citet{hick12}, \citet{mccul14a}, and \citet{fol16}.

In our photometric analysis, we find no stars within 200 parsecs of the SN position. As such, we use the false stars method of {\tt DOLPHOT} to set a lower limit on the delay time for this object. This analysis measures a minimum delay time for SN 2008A of $\rm Age_{08A} \geq 6$ Myr.

\subsection{SN 2008ge}
SN 2008ge was discovered in NGC 1527 \citep{pig08}, at an estimated Tully-Fisher distance of $17.37 \pm 0.96$ Mpc \citep{tully13}. We assume no host galaxy extinction for this object \citep{fol10b}. For this object, we also use a solar metallicity prior \citep{jorg97}.

In our photometric analysis, we find no stars within 200 parsecs of the SN position. As with SN 2008A, we use the false stars method of {\tt DOLPHOT} to set a lower limit on the delay time for this object. From this analysis, we find $\rm Age_{08ge} \geq 18$ Myr.

\subsection{SN 2008ha}
SN 2008ha was discovered in UGC 12682 \citep{puck08}, which has an redshift-derived distance of $21.3 \pm 1.5$ Mpc \citep{fol14}. \citet{lym18} report a metallicity of [Fe/H] $= -0.78 \pm 0.09$ in the region directly around the SN. Additionally, \citet{fol14} report no indication of host extinction in the region around the supernova. For more information on SN 2008ha, see e.g., \citet{val09}, \citet{fol09}, \citet{pum09}, \citet{fry09}, \citet{fol10a}, \citet{stritz14}, and \citet{fol16}. Our analysis uses 16 stars in the vicinity of SN 2008ha to perform the isochrone fitting.

\begin{figure*}
    \includegraphics[width=2.1\columnwidth]{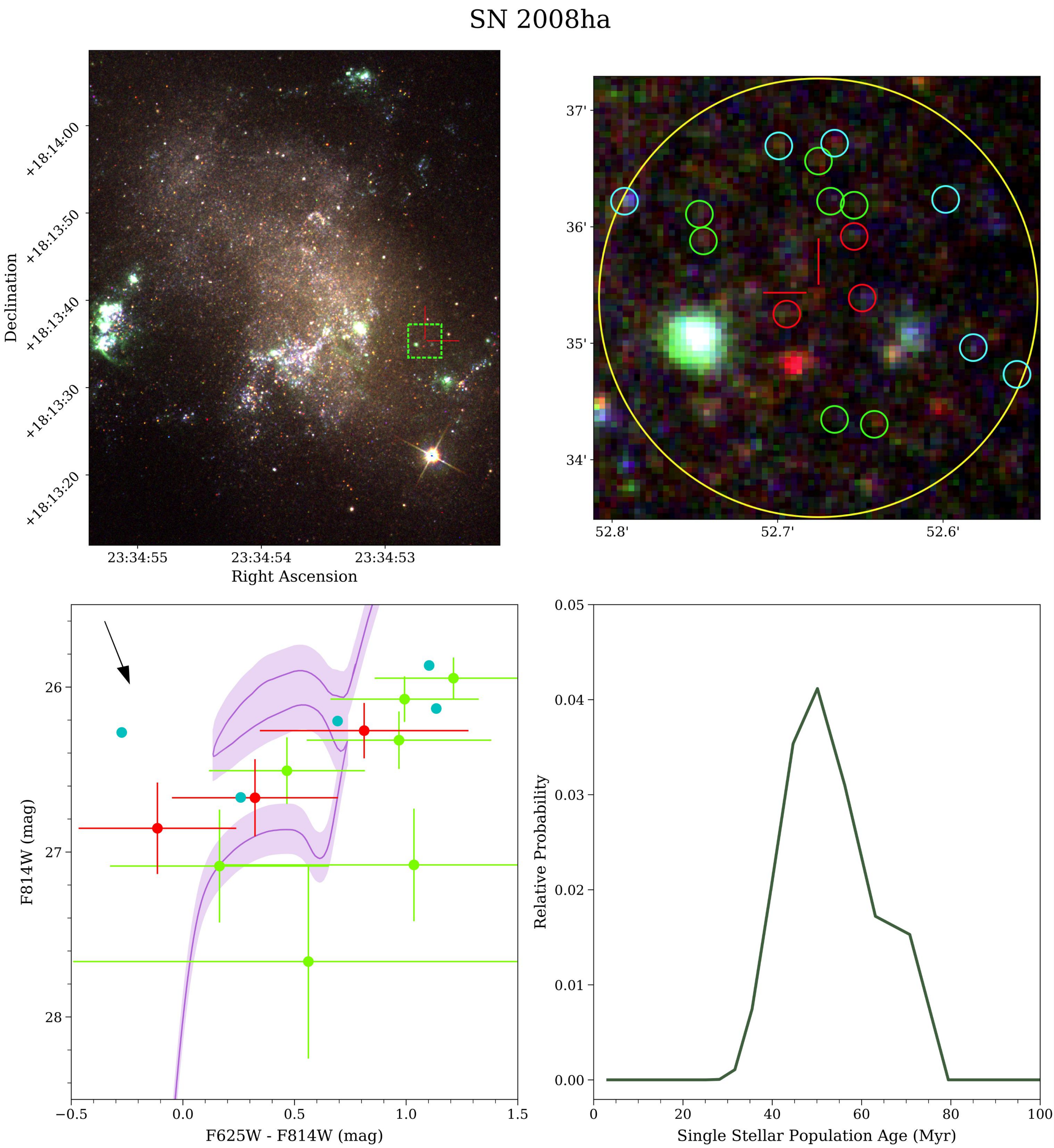}
    \caption{From left to right and top to bottom: a) UGC 12682 with a box with side lengths of 400 pc centered on SN 2008ha. b) The 200 pc radius around SN 2008ha, with the stars used in this study circled. Stars in red are closest to the SN position, followed by those in green, then those in blue. c) A color-magnitude diagram with each of the stars plotted, along with the isochrone for 50 Myr. Error bars are not shown for the blue stars, to avoid overcrowding the diagram. The reddening vector in the upper left shows the direction in which the isochrone would move if there were non-negligible host galaxy extinction. d) The probability distribution for the age of the SN 2008ha progenitor. \label{fig:08ha}}
\end{figure*}

Using the measurement from \citet{lym18} as a prior, metallicity is allowed to float between [Fe/H] = $-0.50$ and $-1.00$. Our fitting algorithm returns a metallicity posterior which is slightly skewed towards solar metallicity, as compared to the prior. Our fits show no preference for any host galaxy reddening for the stars around the SN, consistent with the literature \citep{fol14}. Marginalizing over metallicity, we measure the delay time for SN 2008ha of $52_{- 9}^{+ 13}$ Myr. The full probability distribution in delay time is shown in Figure~\ref{fig:08ha}.

\subsection{SN 2010ae}
SN 2010ae was discovered in ESO 162-17 \citep{pig10}, which is determined using the Tully-Fisher relation to be $11.09 \pm 1.02$ Mpc away \citep{tully13}. \citet{lym18} report a metallicity of [Fe/H] $= -0.43 \pm 0.06$ in the region directly around the SN. A high (though uncertain) upper limit on host extinction of $E (B - V)_{\textrm{host}} = 0.50 \pm 0.42$ mag is reported in the region around the supernova \citep{stritz14}. Additional information on SN 2010ae can be found in e.g., \citet{fol13b} and \citet{fol16}. We find 11 stars in the vicinity of SN 2010ae, which we use for our isochrone fitting.

\begin{figure*}
    \includegraphics[width=2.1\columnwidth]{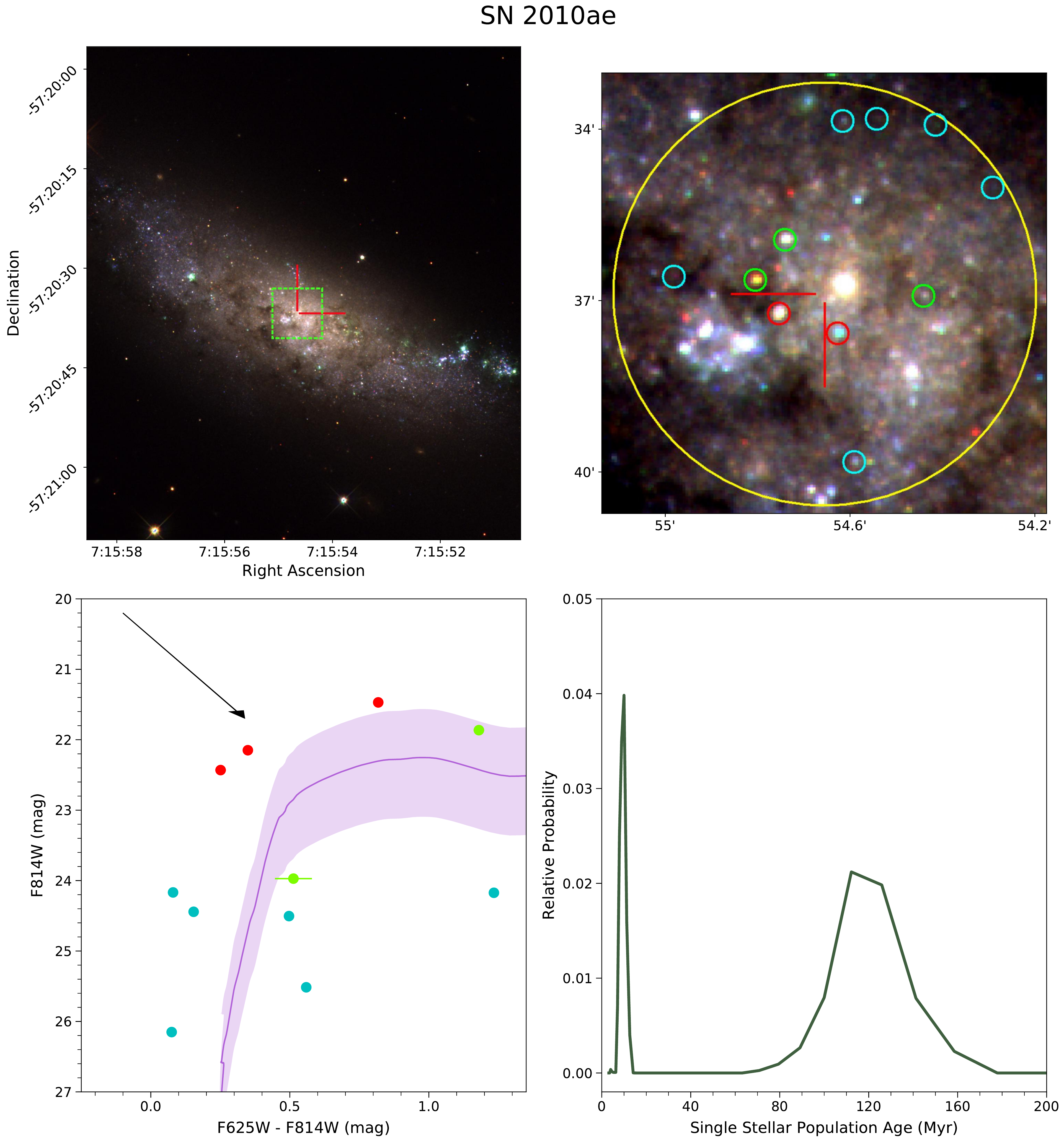}
    \caption{From left to right and top to bottom: a) ESO 162-17 with a box with side lengths of 400 pc centered on SN 2010ae. b) The 200 pc radius around SN 2010ae, with the stars used in this study circled. c) A color-magnitude diagram with each of the stars plotted, along with the median age isochrone of 116 Myr. Stars in red are closest to the SN position, followed by those in green, then those in blue. d) The probability distribution in age for the SN 2010ae progenitor. \label{fig:10ae}}
\end{figure*}

We use Gaussian priors for both metallicity and host-galaxy extinction. The posterior distribution in extinction is skewed towards low extinction values, with a strong preference for $E (B - V)_{\textrm{host}} \leq 0.36$ mag. The posterior distribution in metallicity on the other hand is completely consistent with the prior distribution. Marginalizing over both metallicity and host-galaxy extinction, we find a delay time of $117_{- 29}^{+ 19}$ Myr for SN 2010ae. The full probability distribution for the age of the progenitor is shown in Figure~\ref{fig:10ae}. The measured probability distribution in age has two peaks, one at 9 Myrs, the other at 110 Myrs. This is likely indicative of two separate stellar populations at the explosion site of SN 2010ae, either of which may have been associated with the SN. Providing further evidence for two separate stellar population in the area is the preferred host galaxy extinction values for the isochrone fit. The younger peak favors high values for extinction, while the older peak favors lower values, indicating that the younger population is in the background as compared to the older, as its light passed through more dust.

The probability distribution in delay time that we measure for this object is particularly illuminating, as it demonstrates that our method is able to recover the short delay times that we would expect from a WR progenitor scenario. The young stars around SN 2010ae are likely a stellar population that is unrelated to the SN progenitor, but our detection of this population indicates that our analysis pipeline is working correctly.

\subsection{SN 2010el}
SN 2010el was discovered in NGC 1566 \citep{mon10}, at a Tully-Fisher distance of $5.63 \pm 1.12$ Mpc \citep{tully13}. \citet{lym18} report a metallicity of [Fe/H] $= 0.16 \pm 0.07$ at the site of the SN. No precise measurement is available of the host galaxy extinction in the region around the SN, though a measurement of the extinction of the galaxy at large has been measured to be $E (B - V)_{\textrm{host}} = 0.205$ mag \citep{gouliermis17}. We use 46 stars in the vicinity of SN 2010el to perform our isochrone fitting.

Though we analyse both ACS/WFC data and WFC3/UVIS data for this object, DOLPHOT is better able to identify stars in the ACS/WFC3 data. As such, our ACS/WFC data is far more discernining than the WFC3/UVIS data which we analyze. For this reason, the PDF in age for this object is determined from the ACS/WFC data. All of the analysis for SN 2010el that follows is based on this ACS/WFC data.

\begin{figure*}
    \includegraphics[width=2.1\columnwidth]{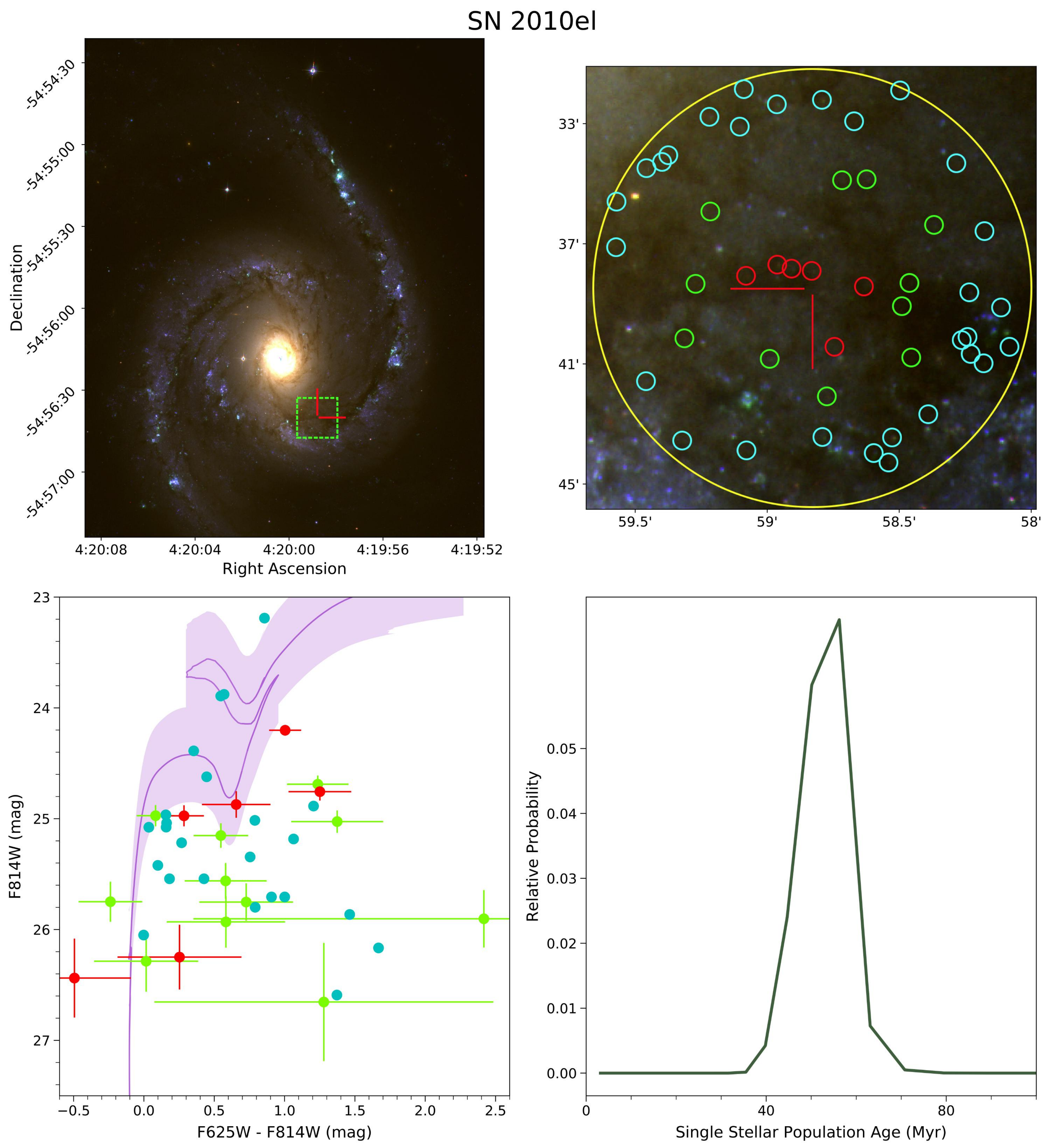}
    \caption{From left to right and top to bottom: a) NGC 1566 with a box with side lengths of 400 pc centered on SN 2010el. b) The 200 pc radius around SN 2010el, with the stars used in this study circled. c) A color-magnitude diagram with each of the stars plotted, along with the best fitting isochrone for 53 Myr. Stars in red are closest to the SN position, followed by those in green, then those in blue. d) The probability distribution in age for the SN 2010el progenitor. \label{fig:10el}}
\end{figure*}

We use the metallicity estimate for the SN as a prior, while assuming a flat prior of $0 \leq E (B - V)_{\textrm{host}} \leq 0.205$ mag for the extinction. Our posterior in metallicity is consistent with the prior, showing a slight preference towards higher metallicity. Our posterior in host galaxy extinction shows a strong preference for non-zero extinction, with the preferred value of $E (B - V)_{\textrm{host}} = 0.205$ mag. When we marginalize over metallicity and host-galaxy extinction, we find a delay time of $53_{- 6}^{+ 5}$ Myr for SN 2010el. The age probability distribution is shown in Figure~\ref{fig:10el}.

\subsection{SN 2011ay}
SN 2011ay was discovered in NGC 2315 \citep{blan11}, to which we measure a distance of $87.4 \pm 6.4$ Mpc, derived from the galaxy's redshift \citep{mill01}. Using the measured metallicity gradient for NGC 2315, we infer a metallicity of [O/H] = $-0.19 \pm 0.20$ \citep{lym18}. The SN has a relatively large offset from its host galaxy, with the SN occurring in a rather empty environment. As such, we assume no host extinction as our prior. For more detailed analysis of SN 2011ay, see \citet{white15}, \citet{sza15}, \citet{fol16}, and \citet{bar17}.

As with SN 2008A and SN 2008ge, in our photometry of NFC 2315, we find no stars within 200 parsecs of the SN location. We perform the same false star tests as with these two objects, in the hope of getting a lower limit on the delay time. However, because of the great distance to NGC 2315, our method cannot rule out any of the isochrones considered in this study.

\subsection{SN 2012Z}
\begin{figure*}
    \includegraphics[width=2.1\columnwidth]{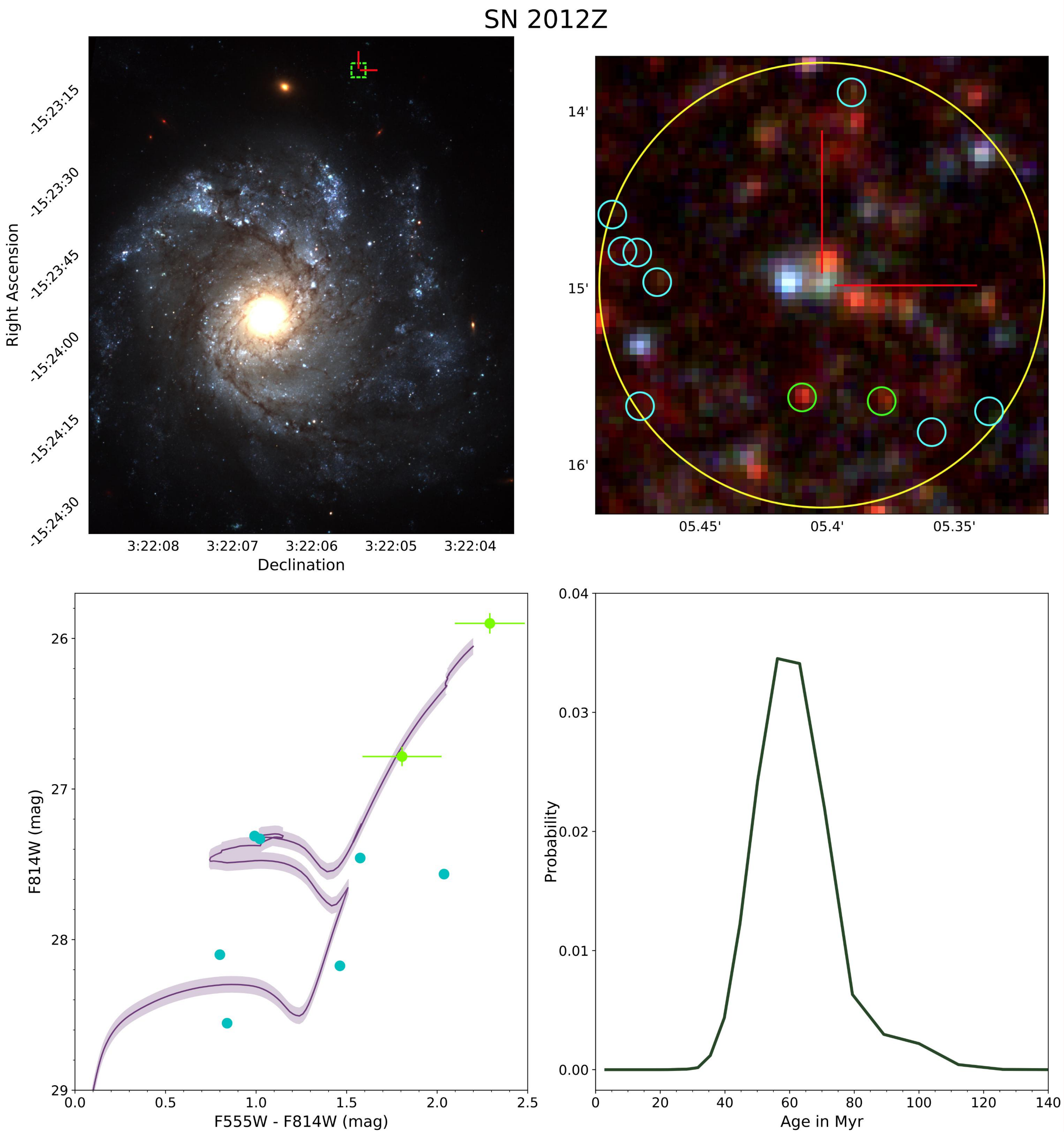}
    \caption{From left to right and top to bottom: a) NGC 1309 with a box with side lengths of 400 pc centered on SN 2012Z. b) The 400 pc box around SN 2012Z, with the stars used in this study circled. c) A color-magnitude diagram with each of the stars plotted, along with the isochrone for 56 Myr. Stars in red are closest to the SN position, followed by those in green, then those in blue. d) The probability distribution in age for the SN 2012Z progenitor. \label{fig:12z}}
\end{figure*}

SN 2012Z was discovered in NGC 1309 \citep{cenk12}, which is determined to be $31.92 \pm 0.88 $ Mpc away \citep{riess16}, using the Cepheid distance method. \citet{lym18} report a metallicity of [Fe/H] $= -0.43 \pm 0.08$ at the site of the SN, in agreement with metallicity gradient that they measure. Host galaxy reddening is estimated  to be $E (B - V)_{\textrm{host}} = 0.07 +- 0.03$ mag using Na I and K I doublets \citep{stritz15}. More information on SN 2012Z can be found in e.g., \citet{mccul14b}, \citet{yama15}, and \citet{fol16}. We use 10 stars in the vicinity of SN 2012Z to perform our isochrone fitting.

Using these measurements as Gaussian priors, we fit for both metallicity and host galaxy extinction. In each case, the calculated posterior is consistent with the prior, showing a slight preference for metallicity and host reddening on the upper end of the ranges given in the literature. Marginalizing over metallicity and extinction, we measure a delay time of $61_{- 11}^{+ 13}$ Myr for SN 2012Z. The full probability distribution in age is shown in Figure~\ref{fig:12z}.

\subsection{SN 2014ck}
\begin{figure*}
    \includegraphics[width=2.1\columnwidth]{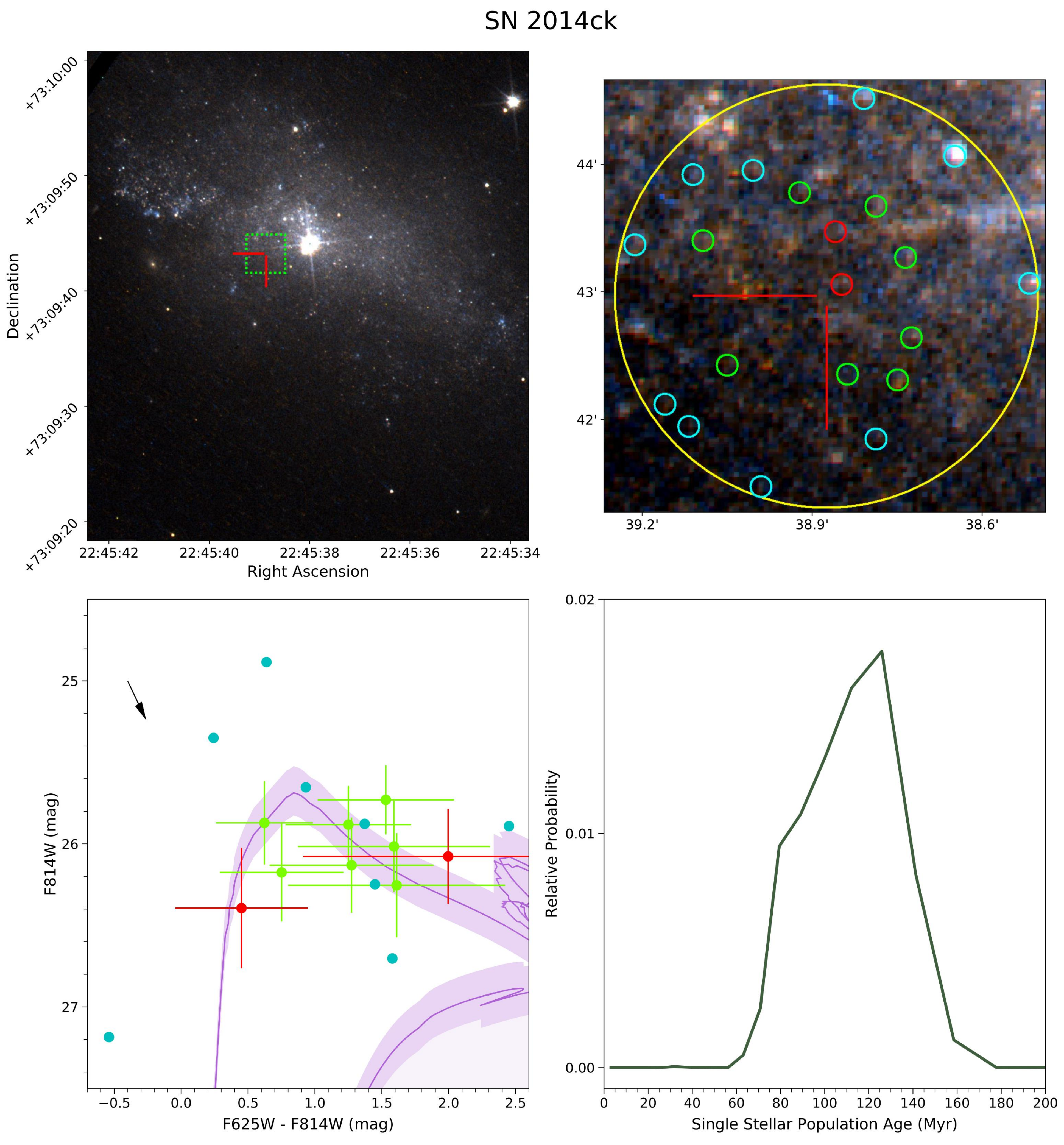}
    \caption{From left to right and top to bottom: a) UGC 12182 with a box with side lengths of 400 pc centered on SN 2014ck. b) The 200 pc radius around SN 2014ck, with the stars used in this study circled. c) A color-magnitude diagram with each of the stars plotted, along with the isochrone for 113 Myr, corresponding to the median of the distribution. Stars in red are closest to the SN position, followed by those in green, then those in blue. d) The probability distribution in age for the SN 2014ck progenitor. \label{fig:14ck}}
\end{figure*}

SN 2014ck was discovered in UGC 12182 \citep{masi14}, which is determined to be $24.32 \pm 1.69$ Mpc away \citep{toma16}, based on its redshift. \citet{tad15} report a metallicity of [Fe/H] $= -0.51 \pm 0.26$ at the site of SN 2006fp, which is also located in UGC 12182, at roughly the same projected distance from the galaxy nucleus. The host galaxy extinction is estimated to be $E (B - V)_{\textrm{host}} \leq 0.05$ mag, determined using the Na I absorption line \citep{toma16}. We use 20 stars in the vicinity of SN 2014ck to perform our isochrone fitting.

These measurements are used as priors, with the metallicity and host galaxy extinction then fit and the posterior in metallicity and extinction measured. Our posterior in metallicity is peaked around the same value as the prior, though with noticeably thinner tails, indicating that our data show a strong preference for [Fe/H] $= -.51$. Our data also indicate no preference for higher extinction, showing consistency with $E (B - V)_{\textrm{host}} = 0$ mag. Marginalizing over both metallicity and extinction, we measure a delay time of $113_{- 25}^{+ 21}$ Myr for SN 2014ck. The full age probability distribution is shown in Figure~\ref{fig:14ck}.

\subsection{SN 2014dt}

SN 2014dt was discovered in Messier 61 \citep{och14} at a distance of $19.3 \pm 0.6$ Mpc \citep{rod14} from our Milky Way galaxy. \citet{lym18} report a metallicity of [Fe/H] $= 0.16 \pm 0.07$ at the explosion site of SN 2014dt, in rough agreement with the metallicity of [Fe/H] $= -0.01$ reported in \citet{fol15}. \citet{fol15} also report no indications of host-galaxy reddening at the site of the SN. We adopt these as priors for our fit.

Our method finds no likely stars within 200 pc of the SN. Performing the false star analysis described in Section 3.3, we are unable to confidently rule out any ages that we test.

\section{Discussion}\label{sec:disc}

\subsection{Verification of Fitting Method}
As mentioned earlier, the probable progenitor star for SN 2012Z has previously been detected \citep{mccul14b}, while a progenitor search for SN 2008ha has yielded strong limits on the delay time for this object \citep{fol14}. In \citet{fol14}, the authors also fit Padova isochrones to the stars directly surrounding the probable remnant of SN 2008ha, to measure an approximate delay-time for the SN. They find a 1-$\sigma$ confidence interval on the age of the system of $55_{- 10}^{+ 13}$ Myr. Using our isochrone fitting method, we find a 1-$\sigma$ confidence interval of $52_{- 9}^{+ 13}$ Myr, in excellent agreement with the literature. In \citet{mccul14b}, the authors similarly fit isochrones by eye to the stars surrounding the possible progenitor of SN 2012Z, deriving a best fitting age range of 10 - 42 Myrs. With our analysis, we find a 1-$\sigma$ confidence internal of $61.0_{- 11}^{+ 13}$ Myr. This disagreement is only at the $\sim 1 \sigma$ level and likely comes from the fact that \citet{mccul14b} used several sources in close proximity to the SN in their fit, which our analysis rejects as being too extended to confidently classify as stars. Our fit instead relies on sources which are further from the SN, but which are more likely to truly be stars. This new preference for higher ages now strongly suggests that SN 2012Z is not consistent with a WR progenitor system.

\subsection{Comparisons Between Objects}
We check for correlations between delay time and peak luminosity, and between delay time and metallicity (as \citep{liu15} suggest). Our peak luminosities are taken from the Open Supernova Catalog \citep{gui17}, while our metallicities are the priors used for our fitting (citations listed above). For this check, we perform Pearson r-tests to measure the strength of a correlation (or anti-correlation) between our measured delay time and either metallicity or peak luminosity. We find r values very close to zero in each case, indicating no statistically significant correlations.

Though SN 2008ge is unique in its host galaxy characteristics, it does not appear to stand out in this study. Instead, we find that the lower limit which we measure from SN 2008ge is completely consistent with the measured DTDs of all but one of the SN in our study. The DTD for SN 2010ae is double peaked, showing two stellar populations. The older of these two populations is consistent with the lower bound from SN 2008ge, though the younger population is not. Because of this broad agreement between SN 2008ge and the sample at large, for the purposes of this study SN 2008ge appears to belong to the class of ``standard'' SNe~Iax. SN 2014ck has also been called an outlier among SNe~Iax due to its observed properties \citep{toma16}. However our analysis does not find that its delay time in significantly different from the delay times of the rest of our sample.

\subsection{SN 2014dt Progenitor Mass Limit}
The false star tests that we perform for SN 2014dt give us a $3-\sigma$ magnitude limit of 24 mag in the F438W filter. Though this is not sufficient to rule out any isochrones that we test, it is sufficient to rule out single stars more massive than $19.5 M_{\odot}$ at the distance of the host galaxy of SN 2014dt. Using MESA single stellar evolution models of the appropriate metallicity, this roughly corresponds to a lower limit on the delay time for this particular object of $10$ Myrs. Because this limit is not derived using the same methods as the rest of the study, we do not include the limit in the rest of the analysis.

\subsection{Empirical Measurements}

\begin{figure}
    \includegraphics[width=\columnwidth]{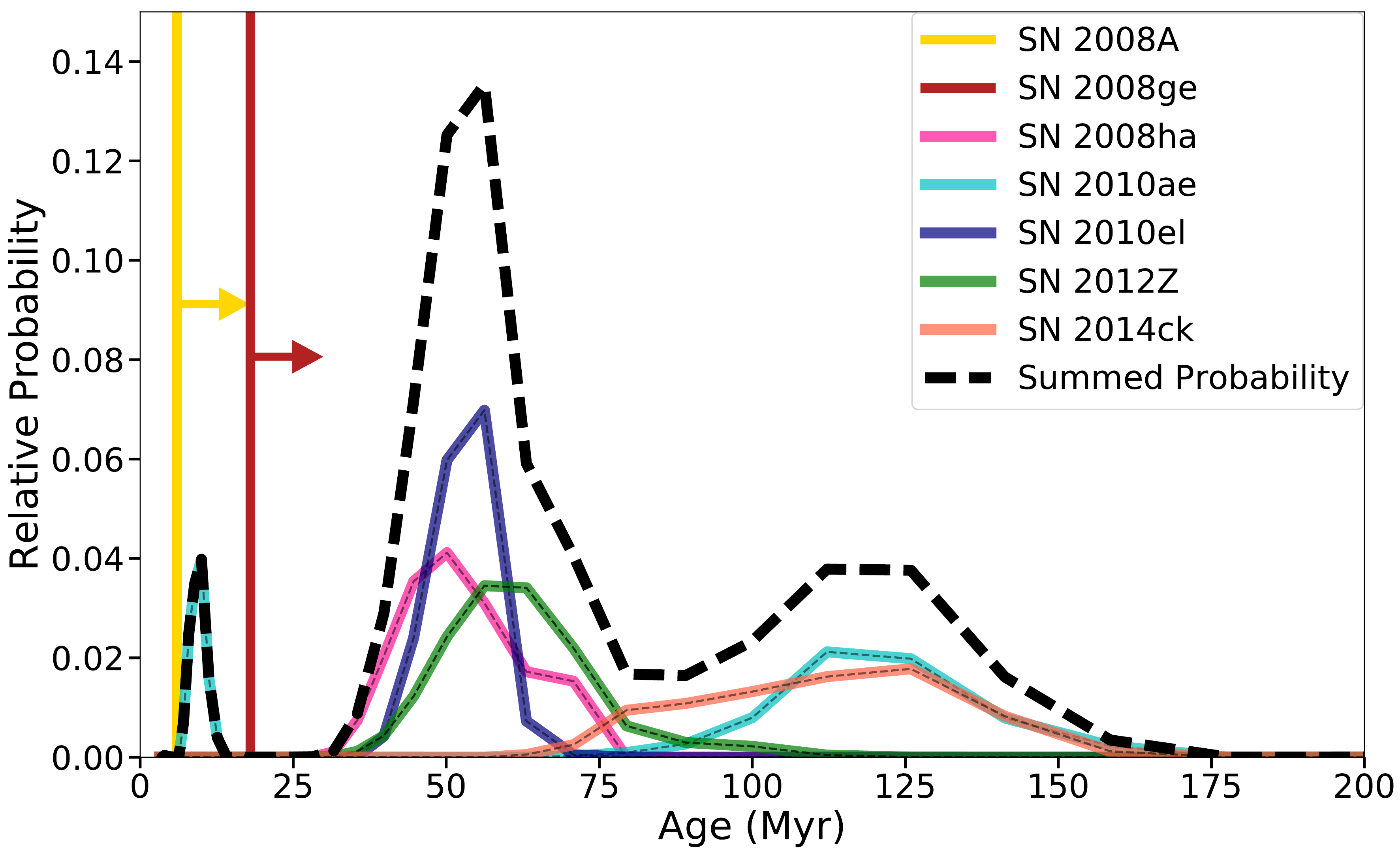}
    \vspace{-0.5cm}
    \caption{The probability density functions (PDFs) for each SN considered in this study. The ``Summed Probability" is the sum of the PDF for each object in the study. As such, it is normalized to 5 (the number of objects in the sample with defined PDFs).\label{fig:pdf}}
\end{figure}

If we assume that all SNe~Iax that we consider in this study share a progenitor channel, we can make strong statistical statements about the delay-time distribution of SNe~Iax at large. The probability density function (PDF) for each SN, and the summed PDF (unnormalized for clarity in the figure) are shown in Figure~\ref{fig:pdf}. Taking only the SNe for which a full DTD can be measured (ignoring lower limits), we find a median delay time and $1-\sigma$ confidence interval of $63_{- 15}^{+ 58}$ Myrs. The data show a strong preference for a delay time peaked near 55 Myrs, with the PDFs for three SNe peaking within 6 Myrs of 55 Myrs. SN 2014ck and SN 2010ae on the other hand, represent the long tail of the distribution to higher ages, with the medians of their DTDs at 113 Myr and 117 Myr respectively.

\begin{figure}
    \includegraphics[width=\columnwidth]{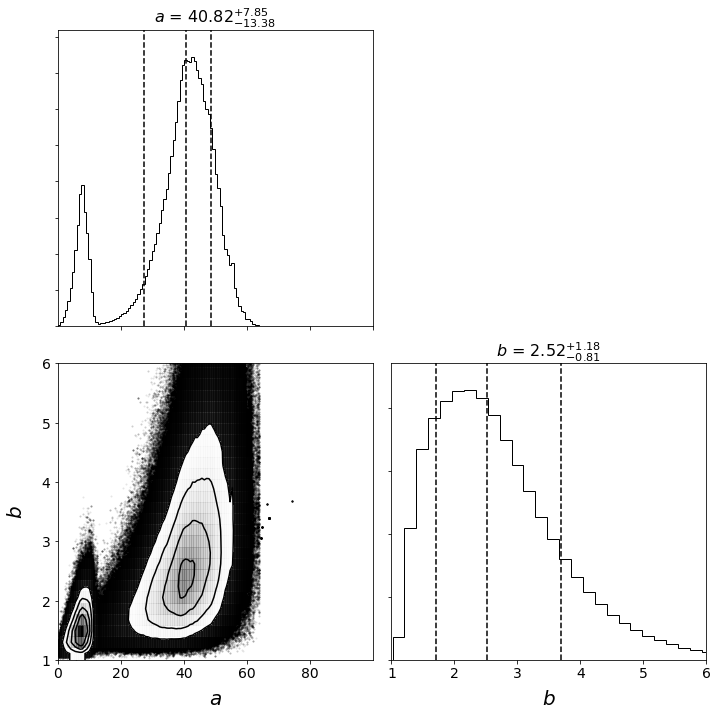}
    \vspace{-0.5cm}
    \caption{The distributions in each parameter used in the power law fit. Parameter $a$ is the cutoff age, while parameter $b$ is the decay power.\label{fig:emcee}}
\end{figure}

If we then draw groups of 5 SNe (the number of SNe~Iax in the summed PDF) from our summed PDF and fit power-law decay models to the result in an MCMC fashion, we find the posterior distribution in Figure~\ref{fig:emcee}. This is done using {\tt emcee} \citep{dfm13} with a flat prior on the cutoff age (0 to 100 Myr) and an inverse-gamma prior with a peak at -1 for the decay exponent. The resulting form of this power-law decay is then described by Equation 2.

\begin{align}
p(t) = \left\{ \begin{array}{cc} 
                0 & \hspace{5mm} t < a \  \text{Myrs}, \\
                 t^{- b} & \hspace{5mm} t >= a \  \text{Myrs}. \\
                \end{array} \right.\label{eqn2}
\end{align}

The distributions of the cutoff age (labeled $a$) and the decay power (labeled $b$) are shown in Figure~\ref{fig:emcee}. The 1-sigma confidence intervals for each parameter are $40.83_{- 13.38}^{+ 7.85}$ Myrs and $2.52_{- 0.81}^{+ 1.18}$ respectively.

As seen in Figure~\ref{fig:emcee}, the distribution in cut-off age ($a$) is double peaked. This double peaked distribution is due to the double peaked measured DTD for SN 2010ae. The less probable, younger peak in $a$ is due to the fit occasionally drawing an age from the younger peak of the SN 2010ae DTD, and trying to fit this alongside the older ages from the other SNe. This could just be the result of a 2nd stellar population forming near the explosion site of SN 2010ae by chance, or it could be indicative of a heterogeneity in SNe formation channels. Repeating this study with a larger sample of SNe~Iax -- for instance, once many more are discovered by the \textit{Large Synoptic Survey Telescope} (LSST) -- would allow for one to answer this question one way or another.

\subsection{Model Comparisons}

In \citet{val09}, \citet{mor10}, and \citet{pum09}, the authors suggest a core-collapse origin for at least some SNe~Iax. These models predict a maximum delay time for SNe~Iax of 6 to 10 Myrs \citep{fry99, heg03}. If we ignore the events with no nearby stars detected, we find just a $2\%$ probability of at least one SN in our sample having a delay time of 10 Myr or less, strongly suggesting a long-lived progenitor channel for SNe~Iax. Using only the analysis of SN 2008ge, we find a $3-\sigma$ minimum age of 18 Myrs, giving us strong evidence that SNe~Iax are not produced by short-lived progenitors.

The WD He-star binary model for the progenitor systems of SNe~Iax instead predicts a delay-time of roughly 30 Myrs and up, with an expected peak in probability at the lower bound delay-time, and a long tail up to much higher ages \citep{wang14, liu15}. The power law decay which we fit to the summed PDF above is a good parameterization of this expected DTD. The best-fit lower limit on delay-time for our data is $40.83$ Myrs, in excellent agreement with the model predictions.

\section{Conclusions}
We have developed a new method to fit simulated photometry of stellar populations generated from {\tt MIST} isochrones to broadband photometric measurements, weighted by distance from a cluster center. We then use this method to fit the stellar neighborhoods of nine SNe~Iax. We employ Bayesian methods to generate probability distributions for the delay times for these nine objects. The probability distributions are shown in Figure~\ref{fig:pdf}. We find a 68\% confidence interval for the delay time of our sample of $63_{- 15}^{+ 58}$ Myrs. When fitting a power law to the overall probability distribution, we find a 68\% confidence interval on the lower bound on the delay time of $40.83_{- 13.38}^{+ 7.85}$ Myrs.

Taken together, this sample of SNe~Iax provides the most precise measurement of the DTD for SNe~Iax to date. The data strongly disfavor a fallback core-collapse origin for SNe~Iax, instead showing consistency with a WD-He star binary progenitor model.

\section*{Acknowledgements}
Based on observations made with the NASA/ESA Hubble Space Telescope, obtained from the Data Archive at the Space Telescope Science Institute, which is operated by the Association of Universities for Research in Astronomy, Inc., under NASA contract NAS 5--26555. These observations are associated with programs GO-12999 and GO-14244.

The authors would like to thank A. Skemer for his helpful advice in the design of the Isochrone Goodness of Fit statistic used in this paper. We also thank J. Maund and C. Kilpatrick for their helpful advice on using {\tt DOLPHOT}. We thank R. Hounsell, and D. Kasen and L. Bildsten for their contributions to one of the two HST proposals that led to this paper. We additionally thank A. Wasserman for his help with the final statistical analyses of the DTD, and J. Schwab for a variety of helpful discussions. Finally, we thank R. Murray-Clay, without whose advice and patience this work would not have been possible.

The UCSC team is supported in part by HST programs GO-12999 and GO-14244, NSF grant AST-1518052, the Gordon \& Betty Moore Foundation, the Heising-Simons Foundation, and by a fellowships from the David and Lucile Packard Foundation to R.J.F. This SN Iax research at Rutgers University is supported by NSF award 161545 and HST programs GO-11133, GO-11590, GO-12913, and GO-13757. M. Stritzinger is supported by a research grant (13261) from the VILLUM FONDEN. 




\bibliographystyle{mnras}
\bibliography{Iax_bib}



\appendix

\section{Additional Information on Isochrone Fitting}\label{app}
The likelihood function which we use in our isochrone fitting scheme is shown below.
\begin{align}\label{eqn1}
\mathrm{Like(Data}, D, R | \mathrm{Iso.}) &= \prod_{i} p(\mathrm{star}_{i}, D, R | \mathrm{Iso.}), \\
         &= \prod_{i} L_{i}, \\
     L_{i} &= \max(L_{ij}), \\
     L_{ij} &= \prod_{k} \frac{1}{\sqrt{2 \pi \sigma_{ik}^2}} e^{-\frac{1}{2 \sigma_{ik}^2} (m_{ik} - m_{\mathrm{Iso}, jk})^2},
\end{align}
where D is distance, R is host galaxy reddening, $i$ specifies a given star, $j$ specifies a given stellar mass in the isochrone, $k$ specifies a given filter, $m_{ik}$ is the magnitude of a given star in a filter, $\sigma_{ik}$ is the corresponding uncertainty, and $m_{\mathrm{Iso}, jk}$ is the magnitude from an isochrone in a given filter for a given mass. $m_{\mathrm{Iso}, jk}$ is the magnitude drawn from the isochrone, adjusted for distance, Milky Way reddening, and host galaxy reddening.

The posterior distribution for an isochrone given a stellar population is then:
\begin{align}
    p(\mathrm{Iso. | Data}) = \iint &p(\mathrm{Iso. | Data}, D, R) \cdot p(D) \cdot p(R) dD dR, \\
    p(\mathrm{Iso. | Data}, D, R) &\propto \mathrm{Like(Data | Iso.}, D, R) \cdot p(\mathrm{Iso.}), \\
    p(\mathrm{Iso.}) &= p(\mathrm{Age}) \cdot p(\mathrm{Metallicity}). \notag
\end{align}
Here $p(\mathrm{Age})$ and $p(\mathrm{Metallicity})$ are the prior distributions in age and metallicity. Our prior in age is a flat distribution between $10^{6.5}$ and $10^{8.5}$ years. Our metallicity priors are described for each SN in section~\ref{sec:indv}. This posterior distribution for an isochrone is the relative probability that the stellar population being considered has a specific age and metallicity, given the stellar data.

Our measured DTD for each SN is this posterior distribution, marginalized over metallicity.
\begin{equation}
    p(\mathrm{Age | Data}) = \int p(\mathrm{Iso. | Data}) p(M) dM,
\end{equation}
where M is metallicity.

\section{Example of Isochrone Fitting}
We will now walk through the process of fitting a set of isochrones to a data set of stars around a supernova. For this example, we will use SN 2008ha. As mentioned in Section 2, we run DOLPHOT on a set of HST images of the host-galaxy of SN 2008ha. We apply the recommended list of cuts on the data output in order to get a list of point sources in the image. This list is restricted to include only those stars which are within 200 pc projected on-sky distance from the SN, assuming a distance to the host-galaxy of $21.3 \pm 1.5$ Mpc. Clusters are removed using the method detailed in Section 2.3 in order to get our final list of stars which we will use for isochrone fitting.

\begin{figure*}
    \includegraphics[width=2.15\columnwidth]{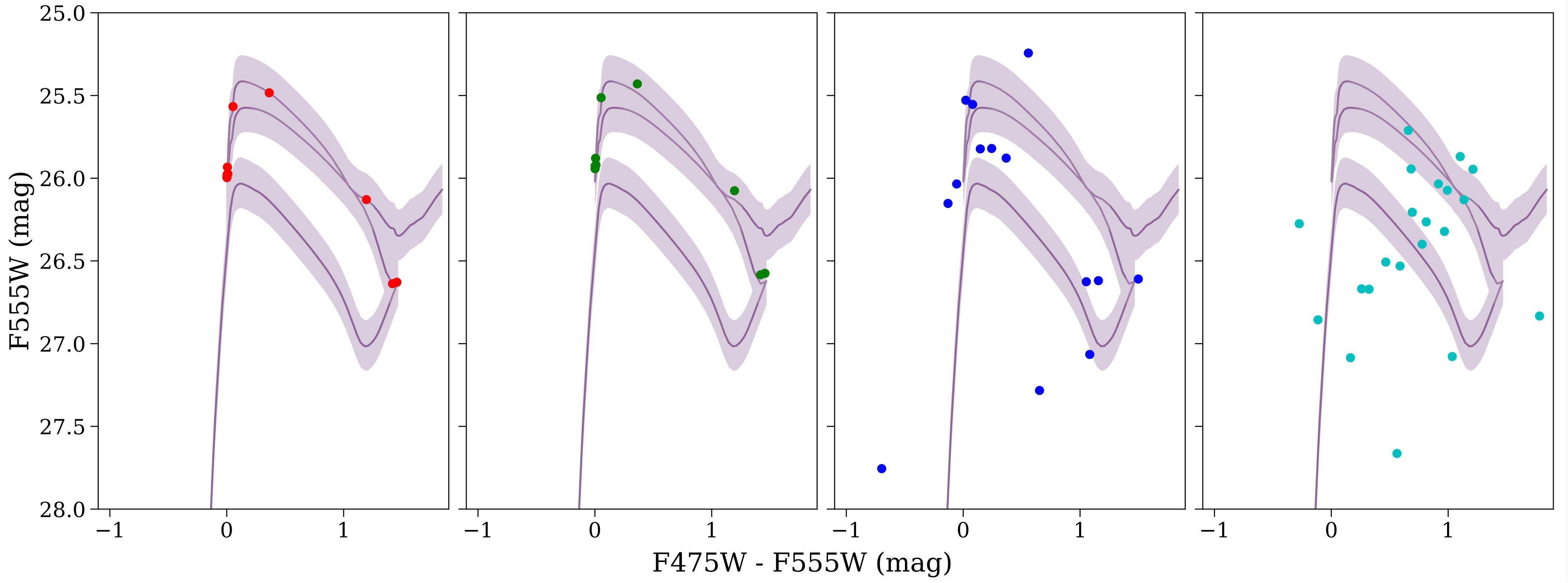}
    \vspace{-0.7cm}
    \caption{From left to right: a) The artificial stars drawn from the $10^{7.45}$ year isochrone with no errors. b) The artificial stars, with a distance error applied. c) The artificial stars, with both the distance error and the magnitude errors applied. d) The real stars for SN 2008ha, indicating general position in Color/Magnitude space. \label{fig:Step}}
    \vspace{-.2cm}
\end{figure*}

The library of isochrones used in this study are logarithmically spaced in age, using 0.05 dex increments from $10^{6.5}$ to $10^{8.5}$ years. These isochrones are drawn using the median, $1-\sigma$, $2-\sigma$ metallicities of our prior, and adjusted for the internal reddening of the host-galaxy. We then construct a distribution to measure probability of a star forming with the SN progenitor, as explained at the beginning of Section 3. Now for each star in our data, we apply the weighting distribution evaluated at the distance of the star. We then find the minimum of the square root of the distance from the star to each isochrone, summed across all of the filters in the data. This process is completed for each star in the data, summing the values measured for each star in quadrature. These values (different for each age, metallicity, and extinction) are our measured Isochrone Goodness of Fit (IGoF) values.

To convert the measured IGoF values to probability distributions in age, metallicity and host-galaxy reddening, we perform careful forward modelling. Stars are drawn from a Kroupa IMF, using the isochrone under consideration to convert the mass to a magnitude in each filter in our data, keeping each star if its associated magnitudes are above our observational limit. Once there are the same number of synthetic stars as real stars in the data, we apply a distance error to all of the synthetic stars together by drawing from a normal distribution with parameters determined by our distance prior. These stars are shown in the second panel of Figure~\ref{fig:Step}. Magnitude errors are applied by drawing from a normal distribution for each star and filter, using the measured magnitude uncertainties as a function of magnitude which are derived from the data. These stars appear in the third panel of Figure~\ref{fig:Step}. We now measure a value of IGoF, and repeat this process 5000 times for each isochrone, in order to get a probability distribution in IGoF for each isochrone. This PDF is used to relate our measured value of IGoF for the isochrone to a relative probability associated with the age, metallicity, and reddening of the isochrone.

The fourth panel of Figure~\ref{fig:Step} shows the real stars detected in the vicinity of SN 2008ha, to compare to the artificial stars which we draw from the isochrone. Note that the real data has higher dispersion than the artificial data because the real data likely features stars which were not born in the same star forming event as the SN progenitor. These stars would have been drawn from isochrones of a different age, and hence would look like a bad fit for the isochrone we plot, whereas every artificial star is drawn from the same isochrone that is plotted.

\begin{figure}
    \includegraphics[width=\columnwidth]{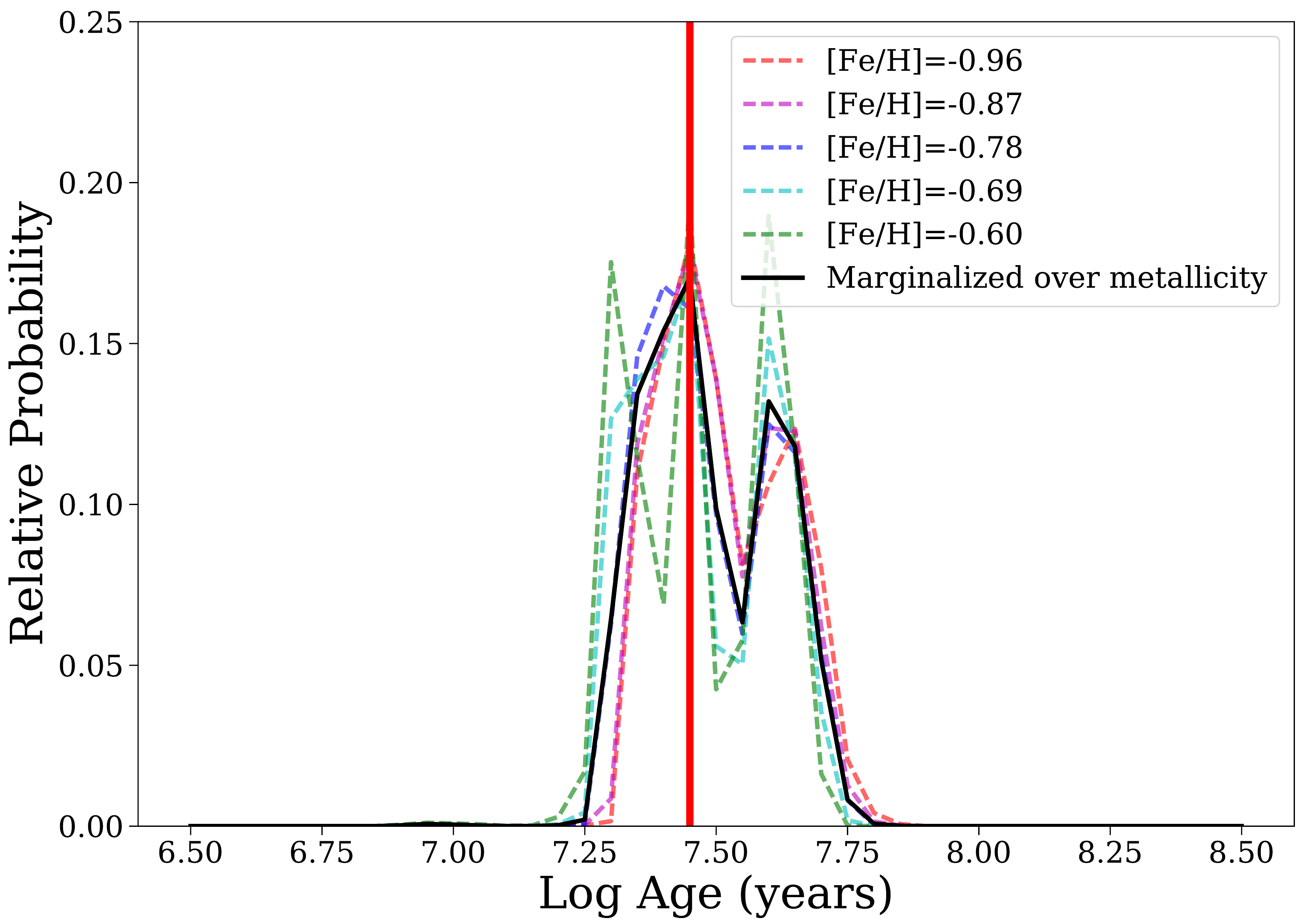}
    \vspace{-.5cm}
    \caption{The probability distribution recovered from an injected age of $10^{7.45}$ years. The injected age is labeled with the vertical red line. The dotted lines show the recovered distributions for a variety of assumed metallicities, while the solid black line shows the distribution from marginalizing over the prior in metallicity. \label{fig:Inj}}
    \vspace{-.3cm}
\end{figure}

\begin{figure}
    \includegraphics[width=\columnwidth]{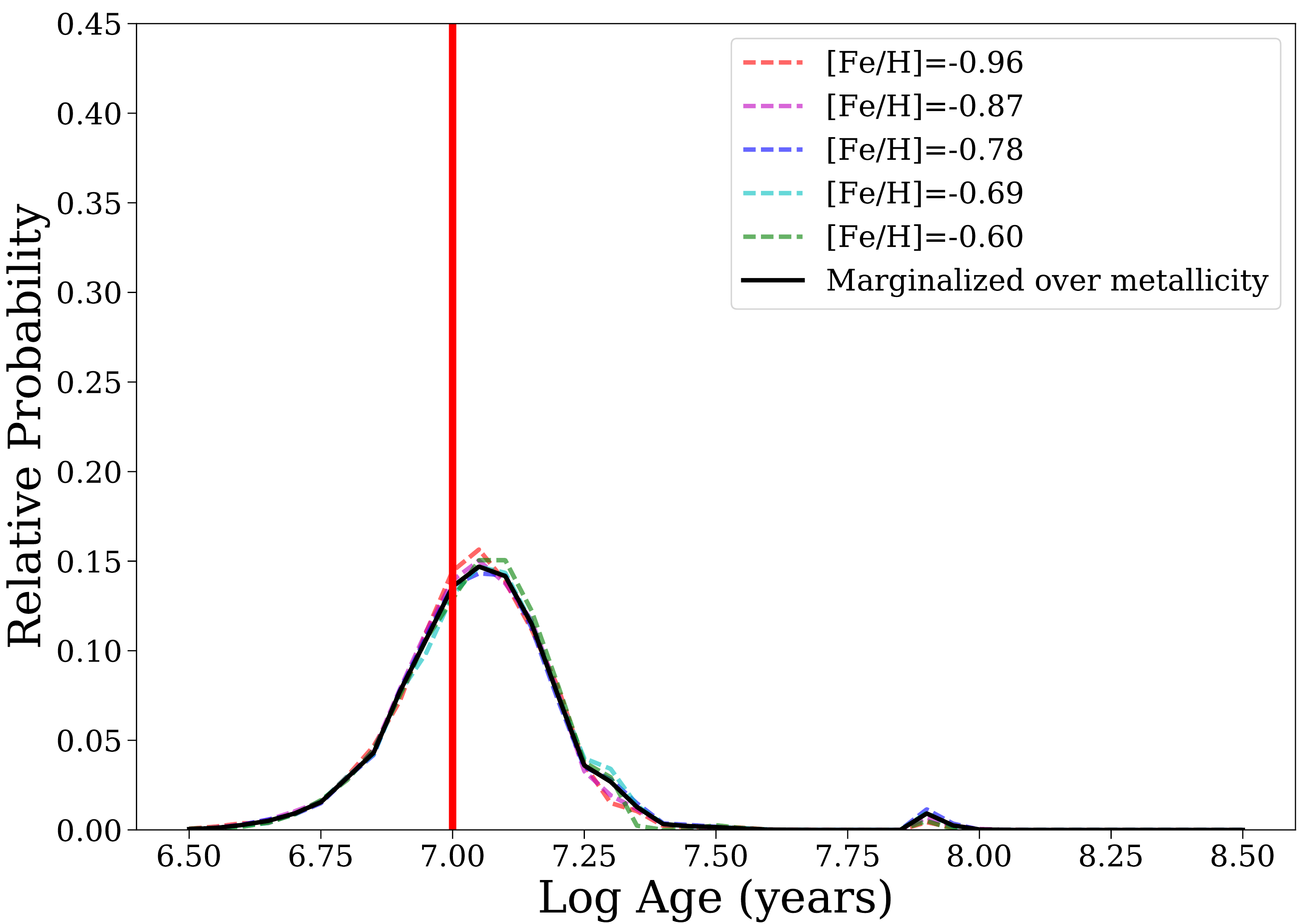}
    \vspace{-.5cm}
    \caption{The probability distribution recovered from an injected age of $10^{7}$ years. The injected age is labeled with the vertical red line. The dotted lines show the recovered distributions for a variety of assumed metallicities, while the solid black line shows the distribution from marginalizing over the prior in metallicity. \label{fig:Inj2}}
    \vspace{-.3cm}
\end{figure}

\begin{figure}
    \includegraphics[width=\columnwidth]{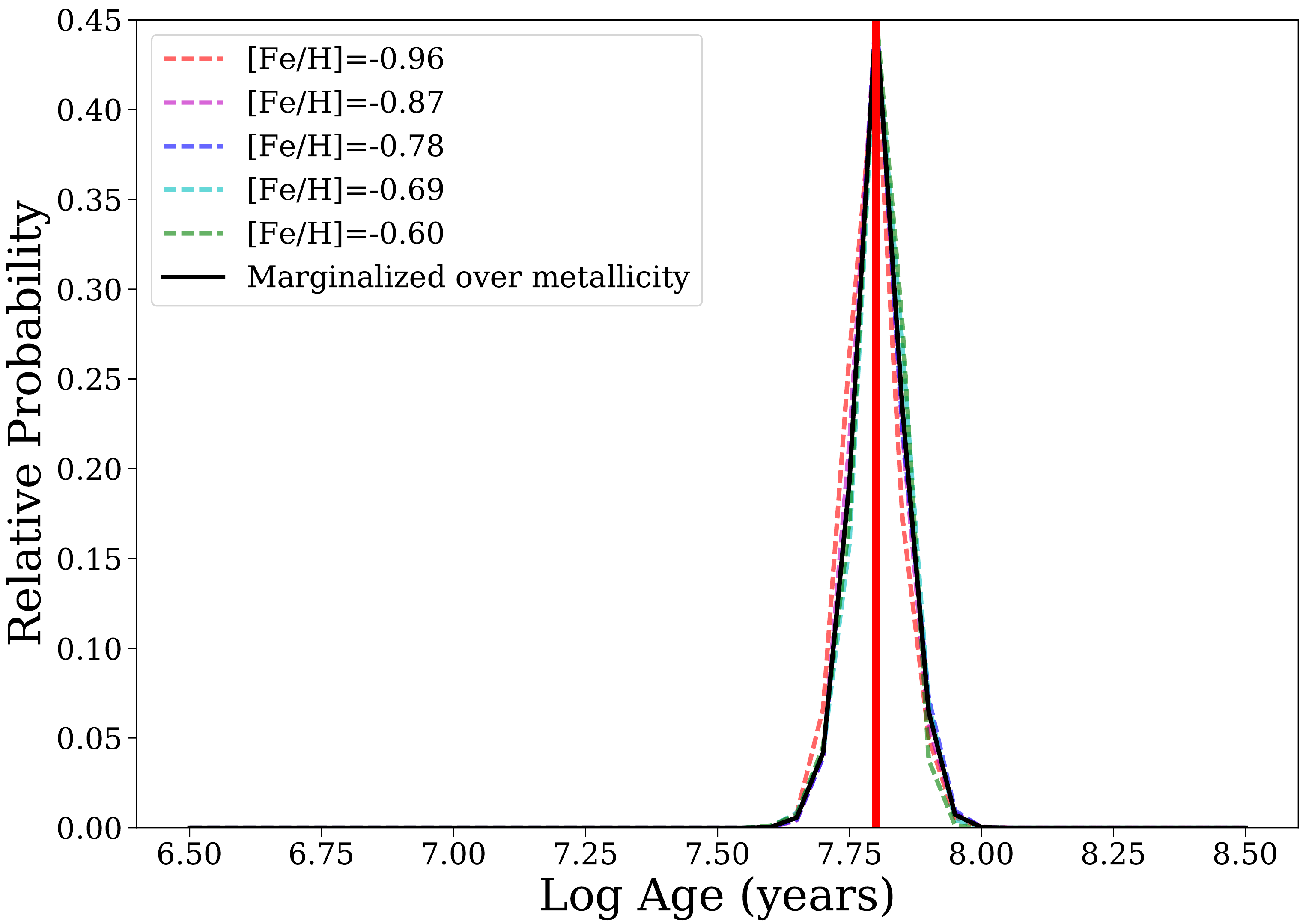}
    \vspace{-.5cm}
    \caption{The probability distribution recovered from an injected age of $10^{7.8}$ years. The injected age is labeled with the vertical red line. The dotted lines show the recovered distributions for a variety of assumed metallicities, while the solid black line shows the distribution from marginalizing over the prior in metallicity. \label{fig:Inj3}}
    \vspace{-.3cm}
\end{figure}

In order to further verify our method, we generate false stars from our isochrone, applying characteristic errors to the stars and our observational sensitivities, and employ our method to recover the age and metallicity of the isochrone. The results of this effort are shown in Figure~\ref{fig:Inj}, Figure~\ref{fig:Inj2}, and Figure~\ref{fig:Inj3}. These figures have injected metallicities that correspond to the median from our prior ([Fe/H] $= -0.78$), while the injected ages are $10^{7.45}$, $10^{7}$, and $10^{7.8}$ years. These tests are successful, as we recover the correct age as the approximate peak in each distribution. Note that each of these tests is performed with 15 stars, in order to match the data for SN 2008ha. For SNe around which we detect more stars (as in SN 2010el, for example), our method recovers the correct age with even higher precision.

\bsp	
\label{lastpage}
\end{document}